\documentclass[9pt]{article}
\usepackage{geometry}
\newgeometry{vmargin={20mm}, hmargin={20mm}}

\usepackage[latin1]{inputenc}
\usepackage[english]{babel}

\usepackage{amsmath,amssymb,stmaryrd,amsthm}
\usepackage{wasysym}
\usepackage{subeqnarray}
\usepackage{verbatim}
\usepackage[colorlinks=false,linktocpage]{hyperref}%
\usepackage{multicol}
\usepackage[bottom]{footmisc}
\usepackage{cancel}
\usepackage{units}
\usepackage{pgf}
\usepackage{afterpage}
\usepackage{gensymb}
\usepackage{cancel}
\usepackage{color,bm}
\usepackage{longtable}
\usepackage[titletoc]{appendix}
\usepackage[nottoc]{tocbibind}
\usepackage{cases}
\usepackage{caption}
\usepackage{enumerate}
\usepackage{graphicx}
\usepackage{subfigure}
\usepackage[square,sort,comma,numbers,super]{natbib}
\usepackage{ulem}

\newcommand{\Eq}{Eq.~}
\newcommand{\Eqs}{Eqs.~}
\newcommand{\Fig}{Fig.~}
\newcommand{\Figs}{Figs.~}
\newcommand{\Tab}{Tab.~}

\newcommand{\rcp}{{\mbox{\scriptsize rcp}}}
\newcommand{\vb}{\mathbf{v}}
\newcommand{\rb}{\mathbf{r}}
\newcommand{\bb}{\mathbf{b}}
\newcommand{\fb}{\mathbf{f}}
\newcommand{\Vb}{\mathbf{V}}

\begin{document}

\begin{center}
\LARGE{Kinetic theory applied to pressure-controlled shear flows of frictionless spheres between rigid, bumpy planes}\\[0.4cm]

\large{Dalila Vescovi,$^{\ast}$\textit{$^{a}$} Astrid S. de Wijn,\textit{$^{b}$} Graham L.W. Cross,\textit{$^{c}$} and Diego Berzi\textit{$^{a}$}}\\[0.4cm]

\normalsize{\textit{$^{a}$~Politecnico di Milano, 20133 Milano, Italy}\\
\textit{$^{b}$~Norwegian University of Science and Technology, NO-7491 Trondheim, Torgarden, Norway.}\\
\textit{$^{c}$~Trinity College Dublin, CRANN, Dublin 2, D02 W085 Ireland.}\\
\textit{$^{\ast}$E-mail: dalila.vescovi@polimi.it}}

\end{center}

\vskip 1 cm

\begin{abstract}
We numerically investigate, through discrete element simulations, the steady flow of identical, frictionless spheres sheared between two parallel, bumpy planes in the absence of gravity and under a fixed normal load. We measure the spatial distributions of solid volume fraction, mean velocity, intensity of agitation and stresses, and confirm previous results on the validity of the equation of state and the viscosity predicted by the kinetic theory of granular gases. In a first, we also directly measure the spatial distributions of the diffusivity and the rate of collisional dissipation of the fluctuation kinetic energy, and successfully test the associated constitutive relations of the kinetic theory. We then phrase and numerically integrate a system of differential equations governing the flow, with suitable boundary conditions, and show a remarkable qualitative and quantitative agreement with the results of the discrete simulations in terms of the dependence of the profiles of the hydrodynamic fields, the ratio of shear stress-to-pressure and the gap between the bumpy planes on the coefficient of collisional restitution, the imposed load and the bumpiness of the planes. Finally, we propose a criterion to predict, on the basis of the solution of the boundary-valued problem, the critical value of the imposed load above which crystallization may occur. This notably reproduces what we observe in the discrete simulations.
\end{abstract} 

\section{Introduction}

In most industrial and geophysical applications involving granular fluids, e.g., hopper discharge, transport on conveyor belts, landslides, the presence of solid boundaries induces strong spatial heterogeneities and controls the flow. 

The discrete nature of granular media suggests the adoption of particle-based numerical methods (discrete element methods, DEM)\cite{cun1979}. However, the high computational costs for solving realistic boundary-valued problems lead to consider continuum approaches as the best suited for dealing with large granular assemblies. Nevertheless, particle simulations may provide physical insights into both the microscopic and the macroscopic dynamics of the granular system, and numerical data can be used to test the continuum models. The latter require appropriate boundary conditions to solve boundary-valued problems and provide the spatial distribution of the associated hydrodynamic fields.

It has been recently shown \cite{ber2024} that the kinetic theory of granular gases \cite{jen1983,lun1991,gar1999,gol2003}, modified to account for correlations in the velocity fluctuations \cite{mit2007}, provides the universal framework to predict the behaviour of granular flows.

The main consequence of the development of velocity correlations is that the rate of collisional dissipation of the fluctuation kinetic energy is over-predicted by classical kinetic theories. This is because the single particle velocity distribution begins to differ from the distribution of the relative velocity between two neighbours \cite{mit2007}, and the measure of the single particle agitation is not anymore indicative of the frequency of collisions. The introduction of a correlation length \cite{jen2006,jen2007,jen2010} in the expression of the rate of dissipation permits the reproduction of the dependence of the granular temperature (one-third of the mean square of the particle velocity fluctuations) on the solid volume fraction in homogeneous shearing flows of rigid spheres\cite{ber2014}.
This extended kinetic theory can successfully reproduce measurements of solid volume fraction, mean velocity, granular temperature and stresses in heterogeneous granular systems, such as gravity driven flows \cite{gol2017,jen2023}, flows through vertical channels \cite{isl2022} and Couette flows \cite{ves2014}, once appropriate boundary conditions are adopted.

In the case of flows over rigid, bumpy planes, boundary conditions for the rate of supply of momentum and fluctuation kinetic energy to nearly elastic, frictionless spheres have been derived by \citet{ric1988}. Empirical modifications for the rate of supply of momentum in the case of inelastic, frictionless spheres have been proposed\cite{ves2014,ber2017}, by using the results of DEM simulations of volume-imposed Couette granular flows.

In this work, we first perform DEM simulations of steady, pressure-imposed, heterogeneous flows of identical, frictionless, inelastic spheres sheared between rigid, parallel bumpy planes, in the absence of gravity, at different values of the particle inelasticity, the imposed pressure and the bumpiness of the planes. We then use the measurements of stresses, solid volume fraction, granular temperature and velocity gradient to assess the validity of the constitutive relations for the particle pressure and viscosity of the kinetic theory, as previously done in different configurations \cite{ves2014,gol2017}. For the first time, to our knowledge, we also test directly the constitutive relations for the flux of fluctuation kinetic energy and for the rate of collisional dissipation. 

Then, we check the validity of the boundary condition for the rate of supply of momentum proposed by \citet{ber2017} and suggest a simple modification to the rate of supply of fluctuation kinetic energy derived by \citet{ric1988} to account for the increase in the energy dissipation for inelastic particles.

We phrase the set of differential equations governing the steady, pressure-imposed shearing flow, using the balance of fluctuation kinetic energy and the constitutive relations of extended kinetic theory, that we numerically solve using the aforementioned boundary conditions, given the total amount of particles in the system. The quantitative agreement between the measurements in the DEM simulations and the predictions of extended kinetic theory are remarkable, in terms of profiles of solid volume fraction, granular temperature, mean velocity and fluctuation energy flux. The theory is also able to notably predict the macroscopic friction (ratio of the particle shear stress to the particle pressure), which represents a measure of the capability of the granular material to act as a lubricant between the moving boundaries, and their gap. 

Finally, we observe in the DEM simulations the transition from a random, fluid state to an ordered, crystalline state when the imposed pressure exceeds a critical value. The extended kinetic theory does not apply in those situations. Nevertheless, by assuming that crystallization occurs when the solid volume fraction at the bumpy boundaries reaches the freezing point, we can fairly predict the dependence of this phase-transition on the particle inelasticity. Although not tested against DEM simulations, we also discuss the dependence of the phase-transition on the bumpiness of the boundaries and the amount of particles in the system.

\section{DEM simulations}\label{Sect_DEM}

\begin{figure}[ht!]
\centering
    \includegraphics[width=16cm]{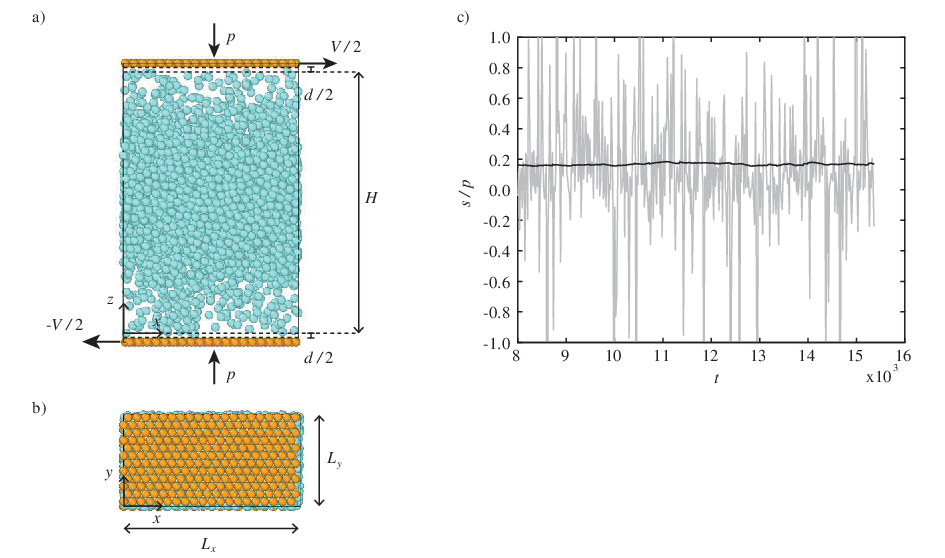}
    \caption{(a) Snapshot of a typical pressure-imposed simulation once the steady state is attained. Also shown is the frame of reference. (b) Top view of the bumpy plane composed of regularly arranged spheres. (c) Temporal evolution of the instantaneous (light grey line) and the smoothed (black line) stress ratio for a typical simulation in steady state.}
    \label{Fig1}
\end{figure}
We study the motion of a collection of identical, frictionless spheres, of diameter $d$ and mass density $\rho_p$, sheared between two parallel bumpy planes, in the absence of gravity (\Fig\ref{Fig1}a). The two bumpy boundaries move at constant relative velocity $V$ in the direction tangent to the planes themselves.
We take $x$ and $z$ to be the flow and shearing directions, respectively, and neglect variations in the transversal direction $y$.  
The planes are made bumpy by gluing spheres of diameter $d_w$ in a regular hexagonal fashion, at close contact and aligned with the direction of the flow, as shown in \Fig\ref{Fig1}(b). We take $z=0$ to be at a distance of $d/2$ from the edges of the bottom boundary particles. The top of the flowing particles is at $z=H$, at a distance of $d/2$ from the edges of the top boundary particles. This choice is motivated by previous studies which looked at boundary conditions for granular flows over rigid bumpy bases \cite{ric1988}.

We use the open-source code LAMMPS \footnote[3]{\href{www.lammps.org}{www.lammps.org}} \cite{LAMMPS} to perform discrete element simulations on a collection of $N = 3150$ spheres interacting via linear spring-dashpots. Periodic boundary conditions are employed in the $x$- and $y$-directions. 
The size of the simulated domain is $L_x = 20$ and $L_y = 6\sqrt{3}$ particle diameters along $x$ and $y$, respectively. The mass of the flowing particles per unit basal area of the rigid boundary is the mass hold-up, $h=\rho_p\left(\pi/6\right)d^3N/\left(L_xL_y\right)$. 
The boundary particles may have a different diameter from the moving particles. 
The bumpiness of the boundaries can be measured through the penetration angle $\psi$, defined as $\sin\psi = (d_w+l)/(d_w+d)$ \cite{ric1988}, where $l$ is the distance between the edges of two adjacent boundary spheres. In this work, as mentioned, we set $l = 0$ and vary $d_w$ in order to investigate the effect of the bumpiness.

In the simulations, all the variables are made dimensionless using particle mass density and diameter and the relative velocity between the boundaries: then e.g., lengths, velocities and stresses are given in units of $d$, $V$ and $\rho_pV^2$, respectively. In the case of frictionless spheres, the collision does not alter the relative velocity in the plane tangent to the point of contact. The particles are characterized by the stiffness $k$ of the linear spring, which is set equal to $2\times 10^5$ for both moving and boundary particles.
The collision between two spheres is completely characterized by the coefficient of normal restitution, $e_n$, that is the negative ratio of the relative velocities of the two colliding particles after and before the collision along the direction normal to the point of contact. In each simulation, we set the same value of $e_n$ for both moving and boundary particles.

The integration time step is set equal to $t_c/50$, where $t_c = \min \left(t_c^{ij}\right)$ and $t_c^{ij}$ is the duration of the collision between particle $i$ and particle $j$ which for the linear spring-dashpot contact model is
\[
t_c^{ij} = \pi \left[ \dfrac{k}{m_{ij}} + \dfrac{k}{m_{ij}}\dfrac{\left(\log e_n\right)^2}{\pi^2+\left(\log e_n\right)^2} \right]^{-1/2},
\]
where $m_{ij} = \pi d_i^3 d_j^3/\left[6 \left(d_i^3 + d_j^3\right)\right]$, being $d_i$ and $d_j$ the diameter of particle $i$ and $j$, respectively. We save data every 5000 simulation time steps, that is the saving time step of the simulations is set equal to 100 $t_c$.

For a given total number of flowing particles in the system, the simulations can be carried out by either (i) imposing the relative distance between the moving planes and measuring the normal force acting on them (volume-imposed simulations) or (ii) imposing the normal force per unit area --the pressure $p$, if we neglect differences in the normal stresses-- acting on the planes and measuring the distance between them (pressure-imposed simulations).

Unlike our previous works \cite{ves2014,ber2017}, here we focus on pressure-imposed conditions, which seem more realistic in view of practical applications of granular materials as lubricants. However, we have performed a large number of discrete simulations and checked that volume-imposed and pressure-imposed simulations would give exactly the same results, if the gap between the planes and the normal force on them are matched.
In order to systematically analyse the role of the coefficient of restitution, the imposed pressure and the bumpiness, we have carried out 25 simulations, with $e_n=\{0.2, 0.4, 0.5, 0.7, 0.8\}$, $p=\{2.9\times 10^{-5}, 2.9\times 10^{-4}, 2.9\times 10^{-3}\}$ and $d_w=\{1/4, 1/2, 1\}$. 

In general, DEM simulations provide the complete microscopic description of the system at each time step, i.e., position and velocity of each particle, as well as  inter-particle forces at contact. Macroscopic, hydrodynamic fields can be determined by appropriate averaging \cite{bab1997b}, here consisting of both spatial and temporal averaging.

The continuum fields, as described in the next Section, are: the $x-$component of the mean velocity, $u$; the solid volume fraction, $\nu$; the pressure, $p$; the shear stress, $s$; the granular temperature, $T$ (one third of the mean square of the particle velocity fluctuations); the rate of collisional dissipation of the fluctuation kinetic energy (the kinetic energy associated with the fluctuations around the particle mean velocity), $\Gamma$; and the $z$-component of the flux of fluctuation kinetic energy, $Q$.

To obtain the local, continuum fields from discrete measurements, we adopt the averaging method described in \cite{bab1997} and summarized in Appendix \ref{appA}, in which the coarse-graining (weighting) function \cite{wei2012,wei2013} is given by the Dirac delta function. We have checked that using a Gaussian rather than a Dirac delta weighting function does not affect the coarse-graining. 

In the absence of gravity, the shear stress and the pressure are spatially uniform. The periodic boundary conditions along $x$ and $y$ ensure that, in the steady state, the remaining variables of the problem can only change along the $z$-direction. We uniformly divide the domain in slices, parallel to the $x$-$y$ plane, of thickness equal to one diameter in the $z$-direction, and we coarse-grain within each slice to obtain local profiles of the continuum variables.

As observed in previous works \cite{mil1996,zam1999}, the measured inter-particle contact forces show large temporal fluctuations, even in the steady state, as contacts are strongly intermittent and the system is still relatively small. 
This is reflected in corresponding large fluctuations of stresses, rate of collisional dissipation and fluctuation energy flux. As an example, \Fig\ref{Fig1}(c) shows the time evolution of the instantaneous stress ratio, $s/p$, in a simulation with $p = 2.9\times 10^{-5}$, $d_w = 1$ and $e_n = 0.5$, in steady state.

The instantaneous continuum fields are smoothed using a moving-average filter. We set the size of the moving-average window equal to 5000 saving time steps, that is the minimum above which the average is independent of the size of the window itself. The smoothed data collected in the steady state, over a total of 20000 saving time steps are then employed to obtain time-averages and standard deviations. This permits to add error bars to the measurement points in the following plots.

\begin{table*}
\centering
\caption{List of auxiliary functions in the constitutive relations of kinetic theory, in which $\Theta\left(\cdot\right)$ is the Heaviside function}
\label{Tab_KTlist}
\begin{tabular}{rl}
\hline
\hline
$J =$ & $\dfrac{1+e_n}{2} + \dfrac{\pi}{32\nu^2 g_0^2} \dfrac{ \left[5+2(1+e_n)(3e_n-1)\nu g_0\right]\left[5+4(1+e_n)\nu g_0\right]}{ 24-6\left(1-e_n\right)^2-5(1-e_n^{2})}$\\[12pt]
$M=$ & $\dfrac{1+e_n}{2} + \dfrac{9\pi}{144\left(1+e_n\right)\nu^2 g_0^2}\dfrac{\left[ 5+3\left(2e_n-1\right)\left(1+e_n\right)^2 \nu g_0\right]\left[5+6\left(1+e_n\right)\nu g_0\right]}{16-7\left(1-e_n\right)}$\\[12pt]
$g_0=$ &  $ \left[1-\Theta\left(\nu-0.4\right)\left(\dfrac{\nu-0.4}{\nu_{rcp}-0.4}\right)^2\right]\dfrac{2-\nu}{2\left(1-\nu\right)^3}+\Theta\left(\nu-0.4\right)\left(\dfrac{\nu-0.4}{\nu_{rcp}-0.4}\right)^2\dfrac{2}{\nu_{rcp}-\nu}$ \\[10pt]
\hline
\hline
\end{tabular}
\end{table*}

\section{Continuum model}

As mentioned, in the absence of gravity, the momentum balances along $x$ and $z$ imply that, in the steady state, the pressure and the shear stress are uniform in the domain. The balance of fluctuation kinetic energy reduces to a balance between diffusion (due to collisions), production (through the work of the shear stress) and dissipation (due to the inelasticity of contacts) \cite{ves2014}

\begin{equation}\label{Eq_EnergyBalance}
Q'= su'-\Gamma.
\end{equation}
Here and in what follows, a prime indicates the derivative with respect to the $z$-direction. \Eq\eqref{Eq_EnergyBalance} is the only balance equation governing the problem, and must be supplied with the constitutive relations for $p$, $s$, $\Gamma$ and $Q$, and appropriate boundary conditions.

\subsection{Constitutive relations}

We adopt the constitutive relations of kinetic theory of granular gases \cite{gar1999}, modified to account for the development of correlations in the velocity fluctuations at solid volume fractions larger than 0.49 \cite{mit2007,jen2007,ber2014}:

\begin{align}
p &= \left[1+2\left(1+e_n\right) \nu g_0\right]\nu  T; \label{Eq_Pressure}\\
s &= \dfrac{8J\nu^2 g_0}{5\pi^{1/2}}  T^{1/2} u'; \label{Eq_ShearStress}\\
\Gamma &= \dfrac{12\left(1-e_n^{2}\right)\nu^2 g_0}{\pi^{1/2}L}T^{3/2}; \label{Eq_Dissipation}\\
Q &= -\dfrac{4 M\nu^2 g_0 }{\pi^{1/2}} T^{1/2} T'. \label{Eq_EnergyFlux}
\end{align}

\noindent Here, the coefficients $J$ and $M$ are explicit functions of the coefficient of normal restitution, the solid volume fraction and the radial distribution function at contact, $g_0$, and are reported in \Tab\ref{Tab_KTlist}.
We employ the expression of the function $g_0$, which accounts for excluded volume and hindering effects and depends on the volume fraction, proposed in Ref.~\cite{ves2014} and reported in \Tab\ref{Tab_KTlist}. This expression combines the well known formulae of Carnahan and  Starling \cite{car1969} and Torquato \cite{tor1995}: like the latter, the radial distribution function of \Tab\ref{Tab_KTlist} diverges when $\nu$ approaches the random close packing, $\nu_\rcp = 0.636$. Unlike the Torquato's expression, however, its derivative with respect to $\nu$ is continuous over the entire range of admissible values of solid volume fraction, thus facilitating the numerical integration of the differential equations.

In \Eq\eqref{Eq_Dissipation}, $L$ is the correlation length of extended kinetic theory \cite{jen2006,jen2007}, a phenomenological quantity which has been introduced to reduce the collisional rate of dissipation whenever correlations in velocity fluctuations are present. Its expression, obtained from discrete numerical simulations of homogeneous shearing, reads \cite{ber2014}:
\begin{equation}\label{Eq_L}
    L = \max\left\{\left[\dfrac{2J}{15\left(1-e_n^2\right)}\right]^{1/2}\left[1+\dfrac{26\left(1-e_n\right)}{15}
    \dfrac{\nu-0.49}{\nu_{rcp}-\nu}\right]^{3/2}\dfrac{u'}{T^{1/2}}, \ 1\right\}.
\end{equation}

The expression for the flux of fluctuation kinetic energy, \Eq\eqref{Eq_EnergyFlux}, should also contain a term proportional to the gradient of the solid volume fraction \cite{gar1999}. However, this term is negligible in dense flows \cite{ber2020}. We will show later that \Eq\eqref{Eq_EnergyFlux} permits the satisfactory reproduction of the measurements in our DEM simulations even in dilute conditions.

We now test the constitutive relations of \Eqs\eqref{Eq_Pressure}--\eqref{Eq_EnergyFlux} against the DEM simulations described in the previous Section. We show the local measurements in five simulations with $e_n = 0.5$ and different values of the imposed pressure and the bumpiness. Similar agreement, not shown here for brevity, is obtained for all the simulations.

\begin{figure}[ht!]
    \centering
    \includegraphics[width=16cm]{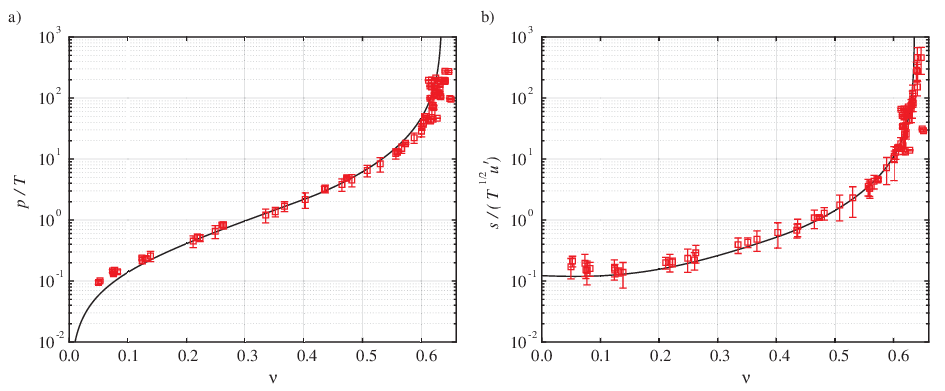}
    \caption{Measurements in DEM simulations (symbols) against predictions of kinetic theory (\Eqs\eqref{Eq_Pressure} and \eqref{Eq_ShearStress}, lines) of (a) scaled pressure and (b) scaled shear viscosity as functions of the solid volume fraction, for $e_n = 0.5$.}
\label{Fig_f1f2}
\end{figure}

\Figs\ref{Fig_f1f2}(a) and (b) confirm that the kinetic theory notably reproduces the behaviour of the measured scaled pressure, $p/T$, and shear stress (shear viscosity), $s/\left(T^{1/2}u'\right)$, over the entire range of solid volume fraction less than the random close packing, even in pressure-imposed heterogeneous shearing flows. The constitutive relations for the pressure and the shear stress have been already tested in previous works on homogeneous shearing flows \cite{ber2015}, volume-imposed heterogeneous shearing flows \cite{ves2014} and inclined flows over rigid, bumpy beds \cite{gol2017}.

\begin{figure}[ht!]
\centering
    \includegraphics[width=16cm]{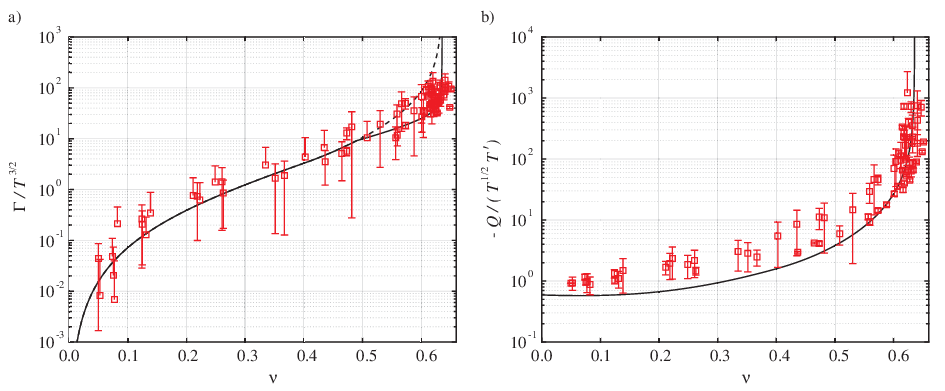}
    \caption{Measurements in DEM simulations (symbols) against predictions of kinetic theory (\Eqs\eqref{Eq_Dissipation} and \eqref{Eq_EnergyFlux}, lines) of (a) the scaled rate of collisional dissipation and (b) the scaled diffusivity as functions of the solid volume fraction, for $e_n = 0.5$. The solid and dashed lines in (a) are \Eq\eqref{Eq_Dissipation} with and without the correlation length $L$ of \Eq\eqref{Eq_Lunif}, respectively.}
    \label{Fig_f3f4}
\end{figure}

Having directly measured in the DEM simulations the rate of collisional dissipation and the flux of fluctuation energy, we are able to test, for the first time to our knowledge, their corresponding constitutive relations. Figure \ref{Fig_f3f4}(a) depicts the scaled rate of collisional dissipation, $\Gamma/T^{3/2}$, as a function of the volume fraction. Once again, the measurements are well reproduced by the expression of kinetic theory, \Eq\eqref{Eq_Dissipation}. At volume fractions larger than 0.49, in the absence of the correlation length, the theory would overestimate the measured rate of collisional dissipation (dashed line in \Fig\ref{Fig_f3f4}a). If $L$ is accounted for, the predictions of the kinetic theory are well within the error bars of the measurements (solid line \Fig\ref{Fig_f3f4}a). For simplicity, in \Fig\ref{Fig_f3f4}a, we employ the expression of the correlation length valid in the case of steady, homogeneous shearing flows \cite{ber2015}, that is when $Q'$ is neglected in \Eq\eqref{Eq_EnergyBalance}:
\begin{equation}\label{Eq_Lunif}
    L = \max\left[ 1+\dfrac{26\left(1-e_n\right)}{15}
    \dfrac{\nu-0.49}{\nu_{rcp}-\nu}, \ 1\right].
\end{equation}
%
Finally, \Fig\ref{Fig_f3f4}(b) shows the dependence of the negative of the scaled flux of fluctuation energy (the scaled pseudo-thermal diffusivity), $-Q/\left(T^{1/2}T'\right)$, on the solid volume fraction, as measured in the DEM simulations and predicted by \Eq\eqref{Eq_EnergyFlux}. The agreement is satisfactory, and suggests that the flux of fluctuation energy mainly depends on the temperature gradient even at solid volume fractions as small as 0.05.

\subsection{System of differential equations and boundary conditions}
Using \Eqs\eqref{Eq_EnergyBalance}--\eqref{Eq_EnergyFlux}, 
and introducing the partial hold-up, $m = \int_0^z \nu d\zeta$ (the particle mass per unit area of the boundary contained below $z$), we phrase the following set of first-order differential equations:\\

\begin{align}
u' &= 5\pi^{1/2}\dfrac{1+2\left(1+e_n\right) \nu g_0}{8J\nu g_0}\dfrac{s}{p}T^{1/2}; \label{Eq_dvxdz}\\
\nu' &= \dfrac{\pi^{1/2}Q}{pT^{1/2}}\dfrac{\left[1+2\left(1+e_n\right) \nu g_0\right]^2}{4 M g_0 }\left[1+2\left(1+e_n\right)\nu \left(2g_0+\nu\dfrac{\partial g_0}{\partial \nu}\right)\right]^{-1}; \label{Eq_dnudz}\\
Q' &= 5\pi^{1/2} p T^{1/2}\left[\dfrac{1+2\left(1+e_n\right) \nu g_0}{8J\nu g_0} \left(\dfrac{s}{p} \right)^2 - \dfrac{1}{5\pi L}\dfrac{12\left(1-e_n^{2}\right)\nu g_0}{1+2\left(1+e_n\right) \nu g_0} \right]; \label{Eq_dqzdz}\\
m' &= \nu. \label{Eq_dmdz}
\end{align}

In \Eq\eqref{Eq_dnudz}, $\partial g_0/\partial\nu$ is the partial derivative of the radial distribution function at contact reported in \Tab\ref{Tab_KTlist} with respect to the solid volume fraction. The correlation length in \Eq\eqref{Eq_dqzdz} is that of \Eq\eqref{Eq_L}, with $u'$ determined from \Eq\eqref{Eq_dvxdz}. The granular temperature is determined algebraically by inverting \Eq\eqref{Eq_Pressure}, as $T=p/\left[\nu+2\left(1+e_n\right) \nu^2 g_0\right]$.

Boundary conditions for collisional flows over rigid, bumpy planes can be expressed in terms of the slip velocity, $u_w$, and the flux of fluctuation energy, $Q_w$, injected into the flow \cite{ric1988}. Then, the set of differential equations, \Eqs\eqref{Eq_dvxdz}--\eqref{Eq_dmdz} is subjected to the following six boundary conditions: 

\begin{align}
m(z = 0) &= 0; \label{Eq_m0}\\
u(z = 0) &= -\dfrac{1}{2} + u_w; \label{Eq_vx0}\\
Q(z = 0) &= Q_w; \label{Eq_qz0} \\
m(z = H) &= h; \label{Eq_mH}\\
u(z = H) &= \dfrac{1}{2} - u_w; \label{Eq_vxH}\\
Q(z = H) &= -Q_w. \label{Eq_qzH}
\end{align}

\noindent The two additional boundary conditions permit the determination of the shear stress $s$ and the height $H$ as parts of the solution.

\begin{figure}[ht!]
\centering
    \includegraphics[width=16cm]{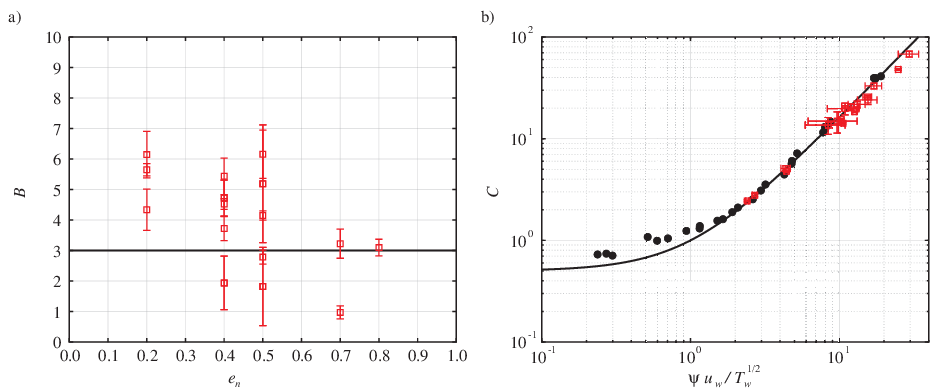}
    \caption{Coefficients (a) $B$ as a function of the coefficient of restitution and (b) $C$ as a function of the scaled slip velocity, measured in DEM simulations of pressure-imposed (open squares) and volume-imposed (filled circles, after Ref.\cite{ber2017}) shearing flows between rigid, bumpy planes. Also shown are the expressions $B=3$ and \Eq\eqref{Eq_C} (solid lines).}
    \label{Fig_BC}
\end{figure}

Using statistical mechanics arguments, Richman \cite{ric1988} derived the flux of fluctuation energy from a rigid, bumpy base in the case of nearly elastic, frictionlesss spheres as
\begin{equation}\label{Eq_EnergyFlux_w}
Q_w = s u_w - \left(\dfrac{2}{\pi}\right)^{1/2}p T_w^{1/2} (1-e_n)  \dfrac{2\left(1-\cos\psi \right)}{\sin^2\psi} B,
\end{equation}
where the first term on the right hand side is the energy produced by the work of the shear stress through the slip velocity, and the second term is the energy dissipated in collisions. There, $T_w$ is the value of the granular temperature at the boundary. Richman \cite{ric1988} obtained $B = 1$ for nearly elastic spheres, but we have found that $B = 3$ permits to better reproduce the results of the DEM simulations, probably because we employ highly dissipative particles. By inverting \Eq\eqref{Eq_EnergyFlux_w}, and using the measurements of $s$, $u_w$, $Q_w$ and $T_w$ from the present DEM simulations permit to plot the coefficient $B$ as a function of the coefficient of restitution, $e_n$ (\Fig\ref{Fig_BC}a). Unfortunately, there is no clear trend, likely because of the difficulties in measuring with sufficient accuracy the energy flux near the boundaries. However, $B=3$ approximates the average of the data inferred from the simulations.

For the slip velocity, a modified version of the the original expression of Richman \cite{ric1988} was proposed by Jenkins et al. \cite{jenUnp},
\begin{equation}\label{SlipVelocity}
\dfrac{u_w}{T_w^{1/2}} = \left(\dfrac{\pi}{2}\right)^{1/2}\dfrac{C}{\psi\csc\psi-\cos\psi}\dfrac{s}{p},
\end{equation}
where $C$ is a function of the scaled slip velocity, $\psi u_w/T_w^{1/2}$. Fitting against DEM simulations of volume-imposed, heterogeneous shearing flows \cite{ber2017} suggested a power-law dependence, with an exponent equal to $3/2$.
This scaling is confirmed in \Fig\ref{Fig_BC}(b), in which we plot the coefficient $C$, obtained by inverting \Eq\eqref{SlipVelocity} and using the measurements of $s$, $u_w$ and $T_w$ in the present DEM simulations, as a function of the measured scaled slip velocity. 
The data can be fitted by the following expression:
\begin{equation}\label{Eq_C}
C = \dfrac{1}{2} + \dfrac{1}{2}\left(\psi\dfrac{u_w}{T_w^{1/2}} \right)^{3/2}.
\end{equation}
We point out that \Eq\eqref{Eq_C} is different from the expression reported in \cite{ber2017}, because the scaled slip velocity was there inadvertently multiplied by the distance $H$.
With this error corrected, the measurements in volume-imposed and pressure-imposed shearing flows between rigid, bumpy planes are in agreement (\Fig\ref{Fig_BC}b).

We solve numerically the system of differential equations \eqref{Eq_dvxdz}--\eqref{Eq_dmdz} with the boundary conditions \eqref{Eq_m0}--\eqref{Eq_qzH}, and \Eqs\eqref{Eq_EnergyFlux_w}--\eqref{Eq_C}, using the 'bvp4c' routine in Matlab. The inputs are the imposed values of $p$, $d_w$ (or alternatively the bumpiness $\psi$), $e_n$ and the mass hold-up $h$. The outputs are the profiles of $u$, $\nu$, $Q$, and $m$, the shear stress $s$ and the distance $H$. From the pressure and the profile of solid volume fraction, the profile of the granular temperature can also be determined from \Eq\eqref{Eq_Pressure}.

\section{Results and comparisons}\label{Results}

In this Section, the predictions provided by the theoretical model are tested against the results of the present DEM simulations. 
We consider separately the effect of (i) the coefficient of restitution, (ii) the imposed pressure and (iii) the bumpiness, on the profiles of mean velocity, solid volume fraction, granular temperature and fluctuation energy flux.

\begin{figure*}
\centering
    \includegraphics[width=16cm]{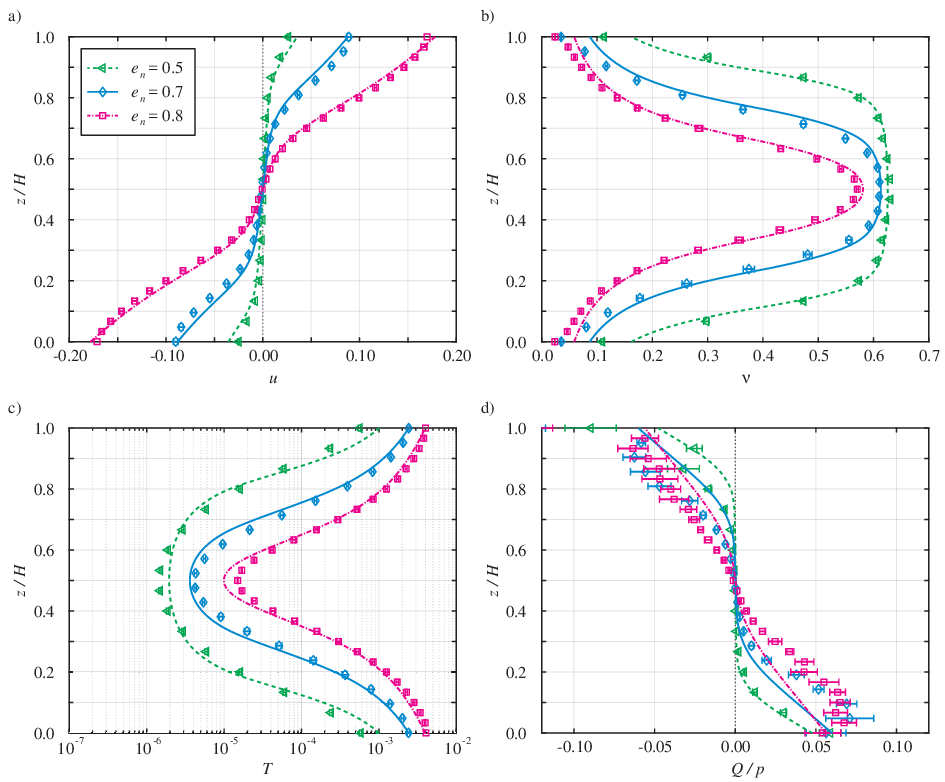}
    \caption{Measured (symbols) and predicted (lines) profiles of (a) mean velocity, (b) solid volume fraction, (c) granular temperature and (d) normalized fluctuation energy flux, $Q/p$, for $p = 2.9\times 10^{-4}$, $d_w = 1$, $h=7.94$, and different values of the coefficient of restitution $e_n$ (see the legend in (a)).}
    \label{Fig_en}
\end{figure*}
In \Fig\ref{Fig_en}, we compare the results of the simulations and the model predictions when the restitution coefficient changes from 0.5 to 0.8, with $p = 2.9\times 10^{-4}$ and $d_w = 1$.
The predictions (lines) are in excellent qualitative and quantitative agreement with the simulations (symbols), for all values of $e_n$. The presence of the bumpy planes induces a strong heterogeneity, with significant spatial gradients in both solid volume fraction and granular temperature. For all values of the coefficient of restitution, the particles form a dense core region, surrounded by two more dilute layers (\Fig\ref{Fig_en}b). This is due to the presence of the bumpy planes, where we observe large slip velocity (\Fig\ref{Fig_en}a) and granular temperature (\Fig\ref{Fig_en}b). Then, the bumpy planes act as local sources of fluctuation energy which originate the two ``hot'' (high $T$) and low density zones, where most of the shear rate is localized (\Fig\ref{Fig_en}a), sandwiching the ``cold'' and dense quiescent core, in which the energy flux vanishes (\Fig\ref{Fig_en}d). This is very much reminiscent of the Leidenfrost effect in molecular fluids \cite{lim2010}.

As already observed in volume-imposed, heterogeneous shearing flows \cite{ber2015},
the size of the dense core and the temperature-difference with the shear zones increase as the coefficient of
restitution decreases. At the smallest restitution coefficient, $e_n = 0.5$, the dense core occupies half of the domain and it is three orders of magnitude colder than the boundaries.

\begin{figure*}
\centering
    \includegraphics[width=16cm]{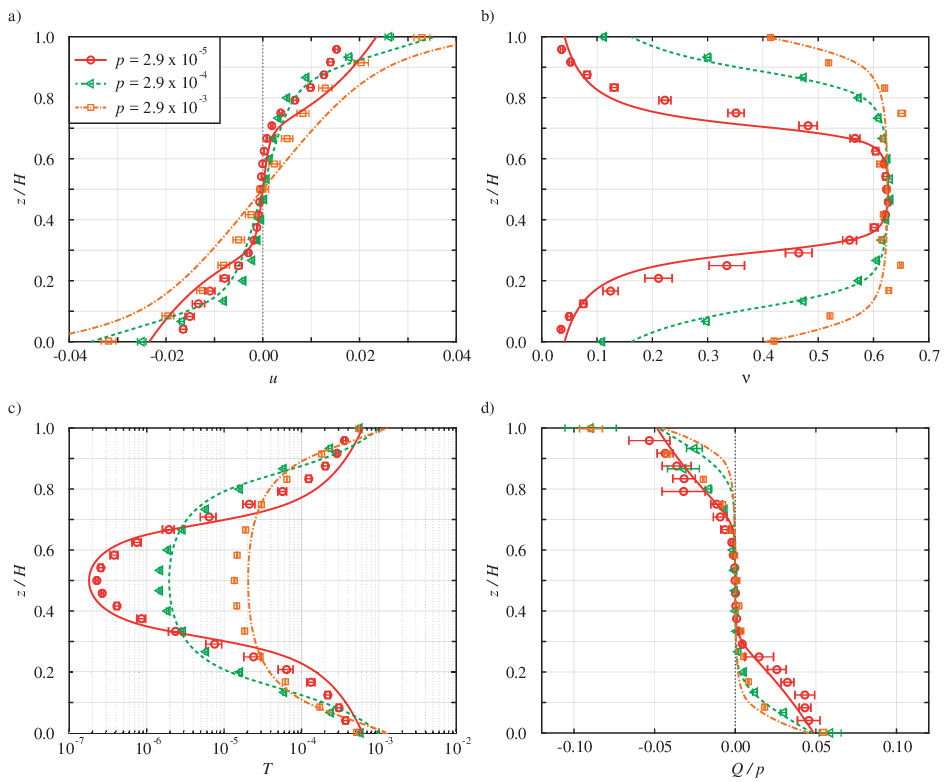}
    \caption{Measured (symbols) and predicted (lines) profiles of (a) mean velocity, (b) solid volume fraction, (c) granular temperature and (d) normalized fluctuation energy flux, $Q/p$, for $e_n = 0.5$, $d_w = 1$,  $h=7.94$, and different values of the imposed pressure $p$ (see the legend in (a)).}
    \label{Fig_p}
\end{figure*}
In \Fig\ref{Fig_p}, we show the effect of varying the imposed pressure over two orders of magnitude when $e_n=0.5$ and $d_w=1$. The pressure has little influence on the profiles of mean velocity (\Fig\ref{Fig_p}a) and normalized flux of fluctuation energy (\Fig\ref{Fig_p}d). Perhaps intuitively, the size of the dense core increases with the pressure, while both the solid volume fraction in the dense core (\Fig\ref{Fig_p}b) and the granular temperature at the boundaries (\Fig\ref{Fig_p}c) are independent of $p$. Conversely, given the relation between $\nu$ and $T$ through the equation of state (\Eq\eqref{Eq_Pressure}), the granular temperature in the dense core and the solid volume fraction at the boundaries strongly increase when the pressure increases. As the pressure increases, the distinction between the dense core and the shear zones fades, and the granular system is much less heterogeneous. We anticipate that the increase of the solid volume fraction at the boundaries with the imposed pressure can explain the transition from a random to a crystalline state that we observe in the DEM simulations for sufficiently high values of $p$. Once again, the continuum model can remarkably reproduce the results of the DEM simulations.

\begin{figure*}
\centering
    \includegraphics[width=16cm]{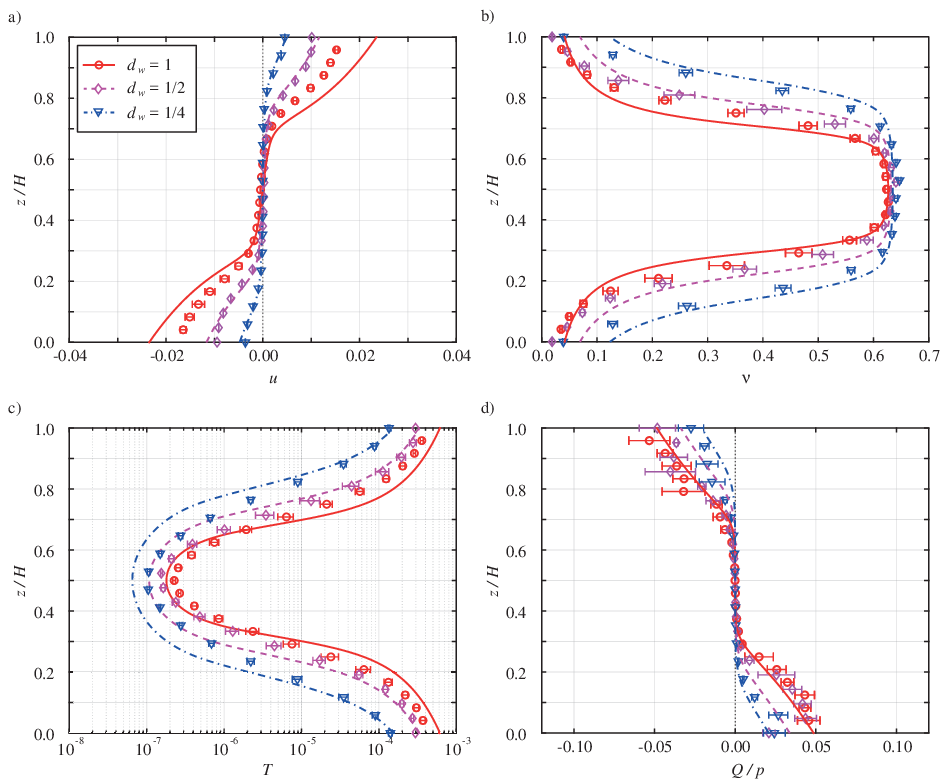}
    \caption{Measured (symbols) and predicted (lines) profiles of (a) mean velocity, (b) solid volume fraction, (c) granular temperature and (d) normalized fluctuation energy flux, $Q/p$, for $e_n = 0.5$, $p = 2.9\times 10^{-5}$,  $h=7.94$, and different values of the boundary particle diameter $d_w$ (see the legend in (a)).}
    \label{Fig_dw}
\end{figure*}
Finally, in \Fig\ref{Fig_dw} we investigate the role of the bumpiness by changing the diameter of the particles which constitute the rigid boundaries, $d_w$, from 1 to 1/4, that is, $\psi$ varies between 0.52 and 0.20, with $e_n=0.5$ and $p=2.9\times 10^{-5}$. The reduction of the bumpiness yields a larger slip velocity at the boundaries (\Fig\ref{Fig_dw}a), and a wider (\Fig\ref{Fig_dw}b) and colder (\Fig\ref{Fig_dw}c) dense core region. This is due to the fact that boundaries that are less bumpy are less effective in transferring particle momentum from the flow, $x$-, to the gradient, $z$-direction, with subsequent reduction in the flux of fluctuation energy from the boundaries to the flow (\Fig\ref{Fig_dw}d). All the profiles are reproduced with high accuracy by the continuum model.

\newpage

\begin{figure}[ht!]
\centering
    \includegraphics[width=16cm]{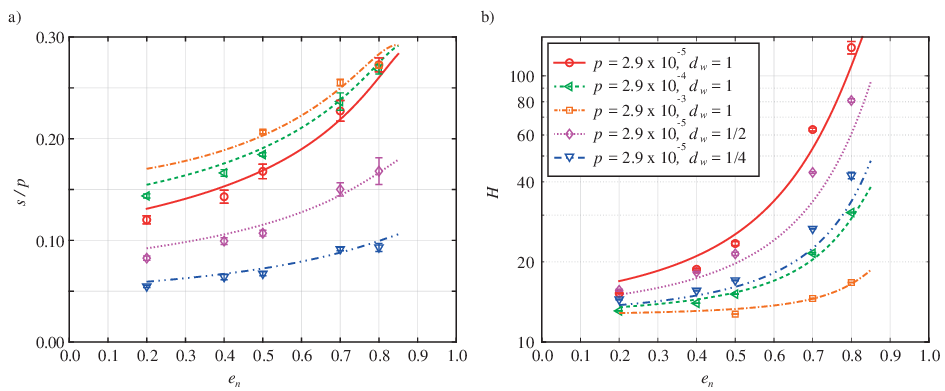}
    \caption{Measured (symbols) and predicted (lines) (a) macroscopic friction and (b) flow height as functions of the coefficient of restitution for $h=7.94$ and: $d_w = 1$ and $p = 2.9\times 10^{-5}$ (red circles); $d_w = 1$ and $p = 2.9\times 10^{-4}$ (green triangles); $d_w = 1$ and $p = 2.9\times 10^{-3}$ (orange squares); $d_w = 1/2$ and $p = 2.9\times 10^{-5}$ (purple diamonds); $d_w = 1/4$ and $p = 2.9\times 10^{-5}$ (blue lower triangles).}
    \label{Fig_mu_h}
\end{figure}

\noindent The ratio of the shear stress to the pressure, $s/p$, is the macroscopic friction of the system and measures the capability of the frictionless moving spheres to act as a lubricant fluid with respect to the two rigid, bumpy boundaries. The dependence of the macroscopic friction on the coefficient of restitution, the imposed pressure and the bumpiness is illustrated in \Fig\ref{Fig_mu_h}(a), where the symbols are the measurements in the DEM simulations and the lines the predictions of the continuum model. The continuum model and the DEM simulations are in remarkable agreement. Similarly, the continuum model correctly predict also the flow height $H$, that is the gap between the rigid, bumpy planes (\Fig\ref{Fig_mu_h}b). 

Counter-intuitively, both the macroscopic friction and the flow height decrease as the particles become more dissipative and the rigid boundaries less bumpy, i.e., $e_n$ and $d_w$ decrease. For instance, we observe a $40\%$ reduction in the macroscopic friction as $d_w$ diminishes from $1$ to $1/4$. Crucially, low values of $s/p$ and $H$ are associated with a strong spatial heterogeneity of the solid volume fraction, and, in particular, with the tendency of the particles to accumulate in a dense, slow-moving core squeezed in between two regions of high shear and high agitation near the bumpy planes.

The imposed pressure plays a minor role in determining the macroscopic friction (\Fig\ref{Fig_mu_h}a). However, the gap between the bumpy planes decreases as the pressure increases by two orders of magnitude (\Fig\ref{Fig_mu_h}b). In this situations, as previously mentioned, the spatial gradients in both $\nu$ and $T$ reduce (\Fig\ref{Fig_p}), and we observe an increase of the macroscopic friction of about $20\%$ (\Fig\ref{Fig_mu_h}a).

\begin{figure}[ht!]
\centering
    \includegraphics[width=8cm]{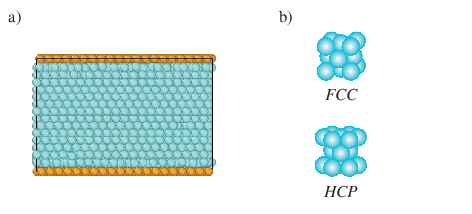}
    \caption{(a) Example of a snapshot taken from the simulation with $e_n = 0.2$, $d_w = 1$ and $p = 2.9\times 10^{-3}$in the steady state, characterized by the presence of crystalline structures inside the flowing particles. (b) Sketch of face centered cubic (FCC) and hexagonal close pack (HCP) structures.}
    \label{Fig_Cryst}
\end{figure}

At large pressure ($p = 2.9\times 10^{-3}$) and small restitution coefficient ($e_n < 0.5$), a persisting steady state cannot be attained even if the total simulation time is increased by one order of magnitude. In such conditions, we observe the formation of regular crystalline structures inside the flowing particles (\Fig\ref{Fig_Cryst}a), that is a dynamic phase-transition from a random to an ordered or partially ordered state.

Measurements of the temporal evolution of the fraction of flowing particles that are arranged in either a face centered cubic (FCC) or a hexagonal close pack (HCP) structure (\Fig\ref{Fig_Cryst}b) during shearing permit to quantitatively identify this phase-transition. Indeed, all the simulations with $p = 2.9\times 10^{-5}$, $p = 2.9\times 10^{-4}$, and those with $p = 2.9\times 10^{-3}$ and $e_n \ge 0.5$, reach a steady state with no flowing particles arranged in regular structures, and such stationary configuration persists for more than 20.000 saving time steps. 
On the other hand, in simulations with $p = 2.9\times 10^{-3}$ and $e_n < 0.5$, we observe a sudden rise of the fraction of flowing particles involved in regular structures, which rapidly reaches a value of about $30\%$.
To identify particles in FCC and HCP structures, we use the adaptive common neighbor analysis as described in \cite{stu2012}, where the standard method \cite{hon1987,fak1994} is extended to multi-phase systems, and implemented in the visualization tool for particle-based systems OVITO\footnote[4]{\href{www.ovito.org}{www.ovito.org}} \cite{stu2010}.

We further investigate the conditions for the phase-transition by performing simulations at imposed pressure larger than $2.9\times10^{-3}$, with various coefficients of restitution, while keeping $d_w = 1$ and $h=7.94$, and measuring the fraction of flowing particles arranged in FCC or HCP structures. We point out that the presence of crystals can be either localized in a small sub-region or span the entire domain. Nevertheless, the fraction of moving particles arranged in regular structures is always larger than $10\%$ whenever crystallization occurs, and even reach $80\%$ at large $p$ and small $e_n$.  

\begin{figure}[ht!]
\centering
    \includegraphics[width=16cm]{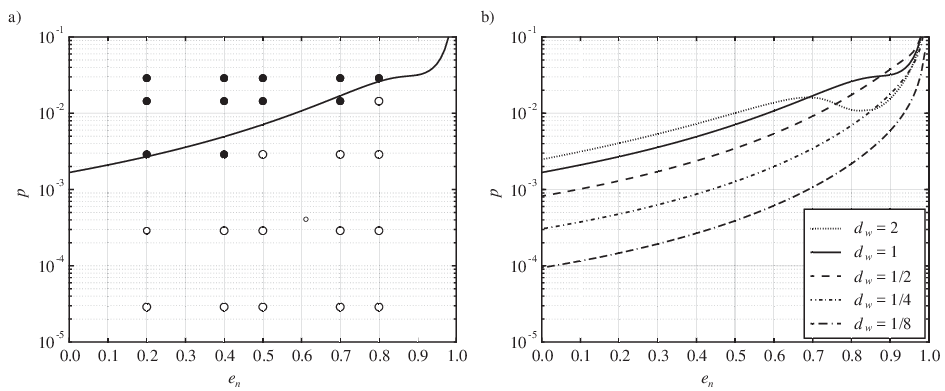}
    \caption{(a) Phase-diagram for the onset of crystallization for $d_w=1$ and $h=7.94$, representing DEM simulations in the presence (filled circles) and in the absence (open circles) of particles arranged in regular structures. The solid line represents the critical pressure predicted from the continuum model. (b) Critical pressure for the onset of crystallization as a function of the coefficient of restitution predicted by the continuum model for $h=7.94$ and different values of the diameter of the boundary particles.}
    \label{Fig_PhaseDiag}
\end{figure}

We can build, then, the phase-diagram depicted in \Fig\ref{Fig_PhaseDiag}(a) which indicates that crystallization occurs when the pressure is larger than a critical value, $p_c$, and that this critical value is an increasing function of the coefficient of restitution. Perhaps intuitively, more dissipative particles crystallize under lower pressures.

Obviously, the continuum model based on the kinetic theory of granular gases, which assumes randomness, cannot apply to crystallized systems. For example, the radial distribution function at contact is singular at the random close packing, while granular solid crystals exist even at larger solid volume fractions. Nonetheless, a phase-transition to an ordered, crystalline state is firstly possible when the solid volume fraction exceeds the freezing point, $\nu=0.49$ \cite{tor1995}. If we assume that the crystal nucleation in the flowing particles is induced by the regular geometry of the bumpy boundaries, we can then estimate that the critical pressure at which crystallization occurs corresponds to the pressure at which the solid volume fraction at the rigid boundaries, $\nu_w$, obtained by solving numerically the differential equations of the continuum model, is equal to 0.49. Operatively, we add the boundary condition $\nu\left(z=0\right)=\nu_w=0.49$ to those of \eqref{Eq_m0}--\eqref{Eq_qzH}, and let $p=p_c$ to be an unknown determined as part of the solution of the system of differential equations \eqref{Eq_dvxdz}--\eqref{Eq_dmdz}. As shown in \Fig\ref{Fig_PhaseDiag}(a), the critical pressure evaluated from the continuum model on the basis of the condition $\nu_w=0.49$ permits to satisfactorily predict the onset of crystallization in the DEM simulations.

\newpage
Interestingly, the critical pressure predicted from the continuum model shows a non-monotonic dependence on the coefficient of restitution for $d_w$ larger than 1, but basically decreases as $d_w$ decreases (\Fig\ref{Fig_PhaseDiag}b). 

\section{Conclusions}

We have performed discrete numerical simulations of steady flows of identical, frictionless, inelastic spheres sheared between parallel, bumpy planes, under pressure-imposed conditions, by systematically varying the coefficient of restitution, the imposed pressure and the bumpiness of the boundaries. 
We have used the results of our simulations to test the predictions of the extended kinetic theory of granular gases.

In particular, we have first validated the constitutive relations of the pressure, the shear viscosity, the rate of collisional dissipation of the fluctuation kinetic energy and the pseudo-thermal diffusivity, for solid volume fractions  smaller than the random close packing.

Then, we have phrased and numerically solved a two-point boundary-valued problem based on the constitutive relations of kinetic theory and the balance of fluctuation kinetic energy, by employing appropriate boundary conditions for the rates of supply of momentum and fluctuation energy into the flow by the bumpy planes. Such boundary conditions have been originally derived by Richman \cite{ric1988} for nearly elastic spheres, by means of statistical mechanics arguments, and later empirically modified by Berzi and Vescovi \cite{ber2017} on the base of numerical simulations. Here we have slightly revised the rate of supply of fluctuation energy to account for the more dissipative particles employed in the present DEM simulations.

We have compared the predicted spatial distributions of mean velocity, solid volume fraction, granular temperature and fluctuation energy flux against the results of the DEM simulations, and have shown remarkable qualitative and quantitative agreement, for all the combinations of input parameters investigated. 
The dependence of the macroscopic friction, i.e., the ratio of the shear stress to the pressure, on the restitution coefficient, the imposed pressure and the bumpiness of the planes is also very well captured by the extended kinetic theory. We have observed that the shear resistance of the flow decreases as the particles become more dissipative and the boundaries less bumpy. As the coefficient of restitution decreases, spatial heterogeneities in the solid volume fraction and granular temperature intensify: the particles tend to accumulate in a dense, slow-moving central core sandwiched in between two more dilute regions of high shear and high agitation near the bumpy planes.

Finally, our simulations have shown that at large imposed pressures, the granular system becomes very dense and  crystallizes. The phase-transition from a disordered/fluid to an ordered/crystalline state takes place at a critical value of the imposed pressure, which depends on the restitution coefficient. The extended kinetic theory can predict the critical pressure at which crystallization occurs by assuming that it corresponds to the pressure at which the solid volume fraction at the bumpy boundaries reaches the freezing point.
The dependence of the critical pressure on the  friction coefficient and a detailed characterization of the behaviour of the granular materials in the sheared crystalline state will be the subject of future works.

\newpage

\section*{Appendix A: Coarse-graining procedure}\label{appA}

Considering each particle $i$ to be a sphere of volume $W_i$, position $\rb^i = \left(r_x^i,r_y^i,r_z^i\right)$ and velocity $\Vb^i = \left(V_x^i,V_y^i,V_z^i\right)$, then volume fraction, $\nu$, flow velocity, $\vb = \left(u,v,w\right)$, and granular temperature, $T$, are computed as

\begin{align}
\nu &= \dfrac{1}{W} \sum_{i} W_i, \label{nu_DEM}\\
u &= \dfrac{\displaystyle\sum_{i} W_i V^i_x}{\displaystyle\sum_{i} W_i}, \  v = \dfrac{\displaystyle\sum_{i} W_i V^i_y}{\displaystyle\sum_{i} W_i}, \  w = \dfrac{\displaystyle\sum_{i} W_i V^i_z}{\displaystyle\sum_{i} W_i} \label{velocity_DEM}\\
T &= \dfrac{1}{3} \dfrac{\displaystyle\sum_{i} W_i \left( \tilde{V}^i_x\tilde{V}^i_x+\tilde{V}^i_y\tilde{V}^i_y+\tilde{V}^i_z\tilde{V}^i_z \right)}{\displaystyle\sum_{i} W_i}, \label{T_DEM}
\end{align}
where $W$ is the local, averaging volume and the summation over $i$ extends over all particles which are wholly contained in $W$.
We have introduced the fluctuation velocity of particle $i$ with respect to the (local) mean velocity: $\tilde{\Vb}^i = \left(\tilde{V}^i_x,\tilde{V}^i_y,\tilde{V}^i_z\right)= \Vb^i-\vb$. \\
Pressure, shear stress and fluctuating energy flux are calculated as the sum of a ``kinetic'' and a ``contact'' contribution:
\begin{align}
p &= p^K + p^C, \label{Pressure_DEM}\\
s &= s^K + s^C, \label{ShearStress_DEM}\\
Q &= Q^K + Q^C. \label{EnergyFlux_DEM}
\end{align}
The kinetic contributions to $p$, $s$ and $Q$ are given as
\begin{align}
p^K &= \dfrac{1}{3} \dfrac{1}{W} \sum_{i} W_i \left( \tilde{V}^i_x\tilde{V}^i_x+\tilde{V}^i_y\tilde{V}^i_y+\tilde{V}^i_z\tilde{V}^i_z \right) , \label{PressureK_DEM}\\
s^K &= \dfrac{1}{W}\sum_{p} W_i \tilde{V}^i_x \tilde{V}^i_z, \label{ShearStressK_DEM}\\
Q^K &= \dfrac{1}{2} \dfrac{1}{W}\sum_{p} W_i \tilde{V}^i_z \left( \tilde{V}^i_x\tilde{V}^i_x+\tilde{V}^i_y\tilde{V}^i_y+\tilde{V}^i_z\tilde{V}^i_z \right), \label{EnergyFluxK_DEM}
\end{align}
whereas contact contributions are
\begin{align}
p^C &= \dfrac{1}{3} \dfrac{1}{W} \sum_{i} \sum_{j > i} \left( f^{ij}_x b^{ij}_x+f^{ij}_y b^{ij}_y+f^{ij}_z b^{ij}_z \right), \label{PressureC_DEM}\\
s^C &= \dfrac{1}{W}\sum_{i}\sum_{j > i} f^{ij}_x b^{ij}_z, \label{ShearStressC_DEM}\\
Q^C &= \dfrac{1}{2}\dfrac{1}{W} \sum_{i}\sum_{j > i} b^{ij}_z [f^{ij}_x \left(\tilde{V}^i_x+ \tilde{V}^j_x\right) 
+ f^{ij}_y \left(\tilde{V}^i_y+ \tilde{V}^j_y\right) \nonumber\\
& + f^{ij}_z \left(\tilde{V}^i_z + \tilde{V}^j_z\right)]. \label{EnergyFluxC_DEM}
\end{align}
In \Eqs\eqref{PressureC_DEM}-\eqref{EnergyFluxC_DEM}, the summation over $i$ and then over $j > i$ extends over all pairs of particles which are in contact and such that the centroids of the contact are located within $W$, counting each pair of particles only once, whereas
$\bb^{ij} = \left(b^i_x,b^i_y,b^i_z\right) = \rb^i - \rb^j$ and $\fb^{ij} = \left(f^{ij}_x,f^{ij}_y,f^{ij}_z\right)$ are the branch vector and the interaction force between particles $i$ and $j$ at contact, respectively.\\
Finally, the rate of dissipation associated to collisions is computed as
\begin{equation}\label{Dissipation_DEM}
\Gamma = \dfrac{1}{W} \sum_{i}\sum_{j > i} \left[f^{ij}_x \left(\tilde{V}^i_x- \tilde{V}^j_x\right) 
+ f^{ij}_y \left(\tilde{V}^i_y- \tilde{V}^j_y\right) + f^{ij}_z \left(\tilde{V}^i_z- \tilde{V}^j_z\right)\right].
\end{equation}

\section*{Author Contributions}

All authors designed research. DV performed simulations and evaluated the data. DV and DB performed the analytical calculations and wrote the manuscript. All authors discussed the results, read, revised and approved the final manuscript.

\section*{Conflicts of interest}
There are no conflicts to declare.

\section*{Acknowledgements}

The authors would like to thank Dr. Eivind Bering for writing the first version of the LAMMPS script.
This project has received funding from the HORIZON-EIC-2021-PATHFINDEROPEN-0 N. 101046693, SSLiP project.
Funded by the European Union. Views and opinions expressed are however those of the author(s) only and do not necessarily reflect those of the European Union or European Commission. Neither the European Union nor the granting authority can be held responsible for them.


\bibliographystyle{plainnat}


\end{document}